# BROAD-BAND BEAM CHOPPER FOR A CW PROTON LINAC AT FERMILAB *

E. Gianfelice-Wendt, V.A. Lebedev[#], N. Solyak, S. Nagaitsev, D. Sun,
Fermilab, Batavia, IL 60510, U.S.A.

*Abstract*

Requirements and technical limitations to the bunch-by-bunch chopper for the Fermilab Project X are discussed.

## REQUIREMENTS

The future Fermilab program in the high energy physics is based on a new facility called the Project X [1] to be built in the following decade. It is based on a 3 MW CW linear accelerator delivering the 3 GeV 1 mA H⁻ beam to a few experiments simultaneously. Small fraction of this beam will be redirected for further acceleration to 8 GeV to be injected to the Recycler/Main Injector for a usage in a neutrino program and other synchrotron based high energy experiments.

After acceleration to 3 GeV the CW beam is RF separated similar to the beam separation in CEBAF where the polarized electron beam is delivered to three experiments simultaneously. At CEBAF the electron bunches are created on the photocathode and the bunch structure for each experiment is set by sequence of laser pulses of corresponding laser. Unlike electron bunches proton bunches cannot be formed at low energy. The beam space charge and much smaller velocity of particles limit the minimum energy where bunches can be manipulated to energies above 1 MeV. Therefore the bunch-by-bunch chopping at the medium energy transport (MEBT) looks as the only option for forming an arbitrary bunch structure. Funnelling beams from two or more RFQs can simplify requirements to chopping but does not eliminate the necessity of bunch-by-bunch chopping in the MEBT. Therefore this choice is not discussed below.

The H- beam is generated in the DC ion source and, then, accelerated in the 2.5 MeV RFQ accelerator. A desired bunch structure is formed by chopping unwanted bunches in the MEBT located between RFQ and SC linac. Chopping reduces beam current. Consequently, the ion source current has to be larger than the beam current in the SC linac. The maximum ion source current of 5 mA is sufficient to support presently considered experiments. Only 1 mA is left after chopping and the rest of the beam should be absorbed on the beam dump.

To mitigate requirements to the chopper kicker the RFQ frequency was chosen to be 162.5 MHz – half of the frequency of the initial part of the SC linac. However even in this case creating required kicks is a challenging problem. It requires chopper parameters well beyond of what is achieved in presently operating systems.

The total beam power coming through the MEBT is 12.5 kW. Even a small beam loss can damage a chopper kicker. A detection of such loss becomes possible if kicker electrodes are not DC coupled to the power amplifier. It also simplifies requirements to the power amplifier. But an absence of DC component requires each kicker pulse to be bipolar. That yields doubling the bandwidth which value should be close to 1 GHz for 162.5 MHz bunch frequency.

## BEAM OPTICS

The low beam velocity and high repetition rate require an electrostatic kicker with wave propagating together with bunch. Then, the kicker strength can be estimated as:

$$eU_k L_k \approx 3 n_\sigma^2 \varepsilon_n \sqrt{mc^2 E_{kin}} \qquad (1)$$

where $\varepsilon_n$ is the rms normalized beam emittance, $e$ and $m$ are the particle charge and mass, $L_k$ is the total kicker length and $E_{kin}$ is the beam kinetic energy. We also assume that the kicker is fed by paraphrase amplifier with voltage $\pm U_k$, the kicker gap and the beam separation on the beam dump are equal to $2n_\sigma \sqrt{\varepsilon \beta_k}$, where $\varepsilon = \varepsilon_n / \beta\gamma$ is the rms emittance, and $\beta_k$ is the beta-function in the kicker center and on the beam dump. CW operation limits the voltage on the kicker to about 300 V. To guarantee good extinction for unwanted bunches the beam separation at the beam dump of 6σ or above is required, consequently $n_\sigma$=3. For $\varepsilon_n$=0.2 mm mrad, $E_{kin}$=2.5 MeV and $U_k$=250 V one obtains $L_k \approx 1$ m. However a single 1 m kicker would require an increase of its aperture due to significant beam displacement at its end. Therefore the kicker is split into blocks separated by 180 deg. in betatron phase. In this case each kicker is short. That also reduces the beam size increase at the kicker ends due to beam divergence. Figure 1 presents the beam envelopes through the MEBT. There are four 50 cm kickers. The first pair excites the beam displacement and the second pair cancels it. Such a scheme allows one not only to chop any bunch but to control each bunch current by its partial scraping. If the kick polarity is changed than the bunch passes the beam dump without scraping. At zero excitation the bunch loses half of its intensity. Twice less kicker voltage is required if all four kickers are used in phase to excite the beam. In this case the unwanted beam is dumped at the line end. To minimize the beam size inside the kickers triplet focusing is used. The longitudinal focusing is performed by 325 MHz RF cavities installed in each second cell. Although strong focusing somewhat reduces harmful effects of the beam space charge the space charge is still quite strong because of small beam emittances. In particular, the beam match into the MEBT performed by the first two triplets and two RF cavities downstream of the RFQ becomes beam current dependent. For the beam current of 5 mA the space charge tune suppressions are: $\Delta v_x/v_x \approx -0.5$, $\Delta v_y/v_y \approx -0.35$ and $\Delta v_s/v_s \approx -0.4$. Simulations of beam propagation through the MEBT showed the emittance growth of ~2% for 5 mA beam. The RF cavities are normal conducting. Their voltage of ~40 kV was chosen to obtain 90 deg. synchrotron phase advance per longitudinal cell (two normal cells). Their transverse defocusing is quite strong and has to be taken into account in optics simulations.

___________________________________________

* Work supported by the U.S. Department of Energy under contract No. DE-AC02-76CH03000
[#]val@fnal.gov

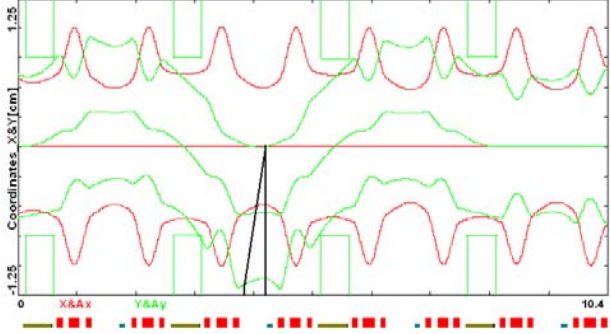

Figure 1. 3σ beam envelopes through MEBT; green line – beam envelope in the plane of kicker excitation, red line – same in the orthogonal plane. Green rectangles and black triangle show positions and apertures of the kickers and the beam dump, respectively. The beam current is 5 mA, $\varepsilon_n$=0.2 mm mrad, $\sigma_s \approx$5 mm, $V_k$= 230 V, the kicker gap is 2·7.4 mm; at the picture bottom: red rectangles show positions of quads, and the blue ones – RF cavities.

## BEAM KICKER

The distance between bunches coming out of RFQ (13.5 cm) is much smaller than the length of one kicker (50 cm). That requires the kicker wave running together with bunch. There are 3 major choices for the kicker design: the meander kicker, the coiled kicker and the kicker based on short kicker plates connected by delay lines. All of them are based on a transmission line where the signal propagation velocity along the beam is adjusted by wiggling the line. Figure 2 presents the geometry of wiggles on a printed circuit board for the CERN meander kicker and a possible geometry for the coiled kicker.

For a straight transmission line the finite resistivity of the conductor and loss in dielectric are the only contributors to the nonlinearity of wave dispersion. For the coiled and meander kickers coupling between lines dominates the dispersion non-linearity. In this case similar to the telegraph equation one writes the equation binding up the voltage and current waves (see Figure 3):

$$\begin{cases} \dfrac{\partial I_n}{\partial x} = -C_0 \dfrac{\partial U_n}{\partial t} + C_0 \kappa_C \left( \dfrac{\partial U_{n+1}}{\partial t} + \dfrac{\partial U_{n-1}}{\partial t} \right) \\ \dfrac{\partial U_n}{\partial x} = -L_0 \dfrac{\partial I_n}{\partial t} \mp L_0 \kappa_L \left( \dfrac{\partial I_{n+1}}{\partial t} + \dfrac{\partial I_{n-1}}{\partial t} \right) \end{cases} \quad (2)$$

Here $n$ numerates the number of line/turn/meander, $C_0$ and $L_0$ are the capacitance and inductance per unit length, $\kappa_C = C_1/C_0$ and $\kappa_L = L_1/L_0$ are the capacitive and inductive coupling coefficients. Sign "-" in Eq. (2) corresponds to the case when currents in nearby lines flow in the same direction (i.e. coiled kicker), and sign "+" otherwise (i.e. meander kicker).

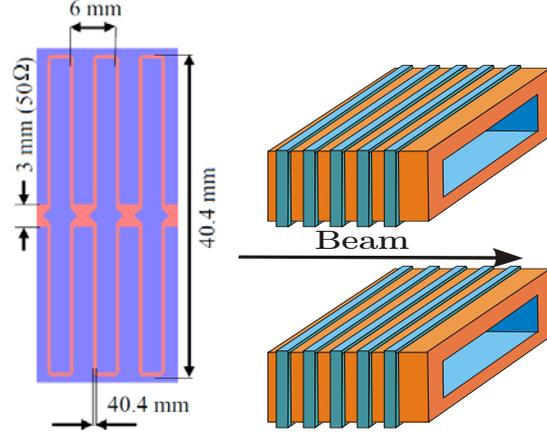

Figure 2. Geometry of printed circuit board for one plate of the CERN meander kicker (left) and geometry and location of electrodes for coiled kicker.

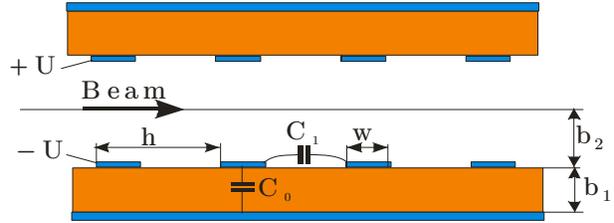

Figure 3. Geometry of coupled transmission lines

In the case of coiled kicker coupling between turns of different coils is much smaller than coupling of nearby turns and Eq. (2) represents good approximation. Solving it with boundary conditions,

$$\begin{cases} U_{n+1}(s,t) = U_n(s+l,t) \\ I_{n+1}(s,t) = I_n(s+l,t) \end{cases} \quad (3)$$

one obtains the following dispersion equation:

$$k \approx \frac{\omega}{v_L}\left[1 - (\kappa_C - \kappa_L)\cos\left(\frac{\omega l}{v_L}\right) - 2\kappa_C \kappa_L \cos^2\left(\frac{\omega l}{v_L}\right)\right], \quad (4)$$

where $v_L \approx 1/\sqrt{C_0 L_0}$ is the wave velocity in the absence of coupling, and $l$ is the one turn length. Similarly for the meander line the boundary conditions are:

$$\begin{cases} U_n(l/2,t) = U_{n+1}(l/2,t) \\ I_n(l/2,t) = I_{n+1}(l/2,t) \end{cases} \quad (5)$$

That yields the dispersion equation:

$$k \approx \frac{\omega}{v_L}\left[1 - (\kappa_C + \kappa_L)\frac{v_L}{2\omega l}\sin\left(\frac{2\omega l}{v_L}\right)\right] \quad (6)$$

where $l$ is the meander width. For $\varepsilon$=1 coupling coefficients are almost equal, $\kappa_C \approx \kappa_L$ and the non-linear contribution to the coiled kicker dispersion (see Eq. (4)) is suppressed. For the meander kicker the contributions from capacitive and inductive couplings are added up but much smaller period length, $l$, moves nonlinearities to higher frequency yielding approximately the same effective dispersion non-linearity and pulse smearing.

For practically interesting cases the strip width is smaller than the dielectric thickness. In this case the capacitance and inductance per unit length are [2] (see Figure 3):

$$C_0(\varepsilon) \approx \frac{1+\varepsilon}{4\ln\left(16 b_1 (1+\varepsilon)/(\pi w \varepsilon)\right)}, \quad w \leq b_1, b_1 \ll b_2. \quad (7)$$
$$L_0 \approx 2\ln(32 b_1/(\pi w))/c^2,$$

Consequently, the coupling coefficients are:

$$\kappa_C \approx \frac{1}{2\ln(16 b_1/\pi w)} \ln\left(\frac{\cosh(\pi h/2 b_1)+1}{\cosh(\pi h/2 b_1)-1}\right), \quad (8)$$
$$\kappa_L \approx \frac{1}{2\ln(32 b_1/\pi w)} \ln\left(1+\left(\frac{2 b_1}{h}\right)^2\right).$$

where we assume that $\varepsilon \gg 1$, $w \leq b_1$, $h \gg w$.

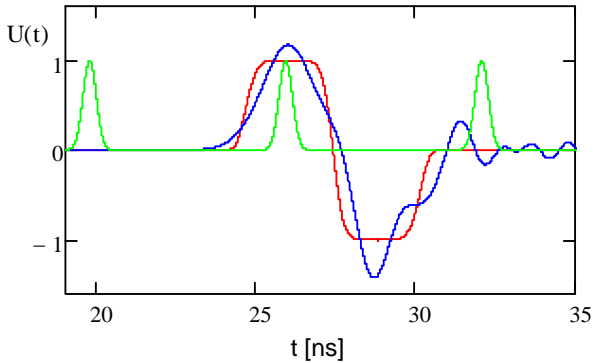

Figure 4. Pulse shape at the kicker entrance (red) and exit (blue). Beam current pulses are shown by green line.

The meander kicker was thoroughly investigated for the SPL [3]. It has the following strip parameters: $\varepsilon$=9.6, $b_1$=3 mm, $b_2$=7 mm, $h$=3 mm and $w$=0.045 mm which yields the coupling coefficients: $\kappa_C \approx 0.12$ and $\kappa_L \approx 0.19$, and that a coupling between strips accelerates the wave propagation by ~1.3 times. Figure 4 presents a distortion of pulse shape after passing a 25 cm kicker. Because of averaging over kicker length the dependence of actual kick on time is somewhere between the initial and final shapes and is at the limit of being acceptable. Although such kicker attracts by simple and straightforward design it has a number of disadvantages. A narrow strip width reduces the kicker efficiency 1.4 times due to large voltage drop near the strip. In addition such kicker has a dielectric surface directly open to the beam. This surface can be charged by the beam resulting in uncontrollable beam kicks.

The coiled kicker has at least a factor of 5 larger one-turn length, $l$, and has also a broken symmetry due to two coils. It yields more nonlinear dispersion than the meander kicker; and, consequently, such a choice is less attractive.

The above consideration points out that an improvement of kicker efficiency requires an increase of the ratio $w/h$ but it results in larger coupling between lines and, consequently, increases the dispersion non-linearity. Therefore a kicker built from short plates connected by delay cables as shown in Figure 5 is presently considered as the preferred choice for the Project X chopper. In this case the overall coupling between lines is reduced because it happens at small relative length. It also allows one to increase the period, $h$, resulting in a reduction of number of electrodes and the dispersion nonlinearity. Such a scheme requires careful match between electrodes and delay lines which has to prevent the wave reflection at transitions from cable to an electrode and back. 3D computer simulations were used for optimizing the electrodes shape and size and the cable-to-electrode transition. For a single electrode the reflection measurements were recently performed and results agree well with simulation. The measurement results are presented in Figure 6. They yield that for 25 cm kicker the reflections make negligible contribution to the pulse distortion. A prototype kicker is presently assembled. The measurements of a complete structure are expected within next few weeks. Electrode-to-electrode coupling is expected to be a major limitation, but in difference to the coiled and meander kickers the combined delay kicker suggests much larger flexibility to address this problem.

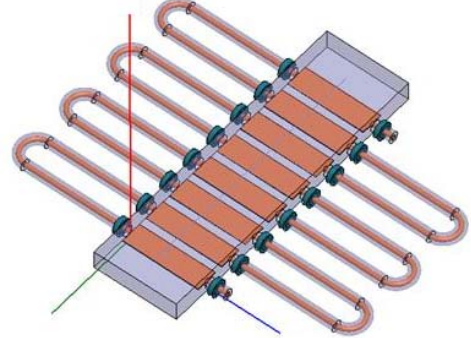

Figure 5. A sketch of combined delay kicker; $L$=40 mm, $w$=18 mm, $h$=23 mm.

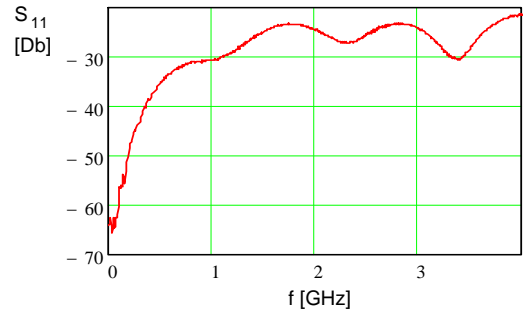

Figure 6. $S_{11}$ for a single electrode.

## REFERENCES

[1] S. Nagaitsev, "Project X – New Multi Megawatt Proton Source in Fermilab", This conference.
[2] Gelmont, et.al., Solid-State Electronics, vol. 38, **3**, pp. 731-734 (1995).
[3] T. Kroyer, F. Caspers, E. Mahner, CARE-Report-06-033-HIPPI (2006).